# Coarse-grained cellular automaton simulation of spherulite growth during polymer crystallization

Dierk Raabe

Max–Planck–Institut für Eisenforschung,

Max–Planck–Str. 1, 40237 Düsseldorf, Germany (d.raabe@mpie.de)

*keywords: texture, polymer, crystallization, spherulite growth, microstructure, transformation, Avrami*

## Abstract

The work introduces a 3D cellular automaton model for the spatial and crystallographic prediction of spherulite growth phenomena in polymers at the mesoscopic scale. The automaton is discrete in time, real space, and orientation space. The kinetics is formulated according to the Hoffman-Lauritzen secondary surface nucleation and growth theory for spherulite expansion. It is used to calculate the switching probability of each grid point as a function of its previous state and the state of the neighboring grid points. The actual switching decision is made by evaluating the local switching probability using a Monte Carlo step. The growth rule is scaled by the ratio of the local and the maximum interface energies, the local and maximum occurring Gibbs free energy of transformation, the local and maximum occurring temperature, and by the spacing of the grid points. The use of experimental input data provides a real time and space scale.

# 1 Introduction to cellular automaton modeling of spherulite growth

## 1.1 Basics of cellular microstructure automata

Cellular automata are algorithms that describe the spatial and temporal evolution of complex systems by applying local switching rules to the discrete cells of a regular lattice [1]. Each cell is characterized in terms of state variables which assume one out of a finite set of states (such as crystalline or amorphous), but continuous variable states are admissible as well (e.g. crystal orientation) [2]. The opening state of an automaton is defined by mapping the initial distribution of the values of the chosen state variables onto the lattice. The dynamical evolution of the automaton takes place through the synchronous application of switching rules acting on the state of each cell. These rules determine the state of a lattice point as a function of its previous state and the state of the neighboring sites. The number, arrangement, and range of the neighbor sites used by the switching rule for calculating a state switch determines the range of the interaction and the local shape of the areas which evolve. After each discrete time interval the values of the state variables are updated for all lattice points in synchrony mapping the new (or unchanged) values assigned to them through the local rules. Owing to these features, cellular automata provide a discrete method of simulating the evolution of complex dynamical systems which contain large numbers of similar components on the basis of their local interactions. The basic rational of cellular automata is to try to describe the evolution of complex systems by simulating them on the basis of the elementary dynamics of the interacting constituents following simple generic rules. In other words the cellular automaton approach pursues the goal to let the *global* complexity of dynamical systems emerge by the repeated interaction of *local* rules.

The cellular automaton method presented in this study is a tool for predicting microstructure, kinetics, and texture of crystallizing polymers. It is formulated on the basis of a scaled version of the Hoffman-Lauritzen rate theory for spherulite growth in partially crystalline polymers and tracks the growth and impingement of such spheres at the mesoscopic scale on a real time basis. The major advantage of a discrete kinetic mesoscale method for the simulation of polymer crystallization is that it considers material heterogeneity [2] in terms of crystallography, energy, and temperature, as opposed to classical statistical approaches which are based on the assumption of material homogeneity.

## 1.2    Cellular automaton models in physical metallurgy

Areas where microstructure based cellular automaton models have been successfully introduced are primarily in the domain of physical metallurgy. Important examples are static primary recrystallization and recovery [2-9], formation of dendritic grain structures in solidification processes [10-13], as well as related nucleation and coarsening phenomena [14-16]. Overviews on cellular automata for materials and metallurgical applications are given in [2] and [17].

## 1.3    Previous models for polymer crystallization

Earlier approaches to the modeling of crystallization processes in polymers were suggested and discussed in detail by various groups. For instance Koscher and Fulchiron [18] studied the influence of shear on polypropylene crystallization experimentally and in terms of a kinetic model for crystallization under quiescent conditions. Their approach was based on a classical topological Avrami-Johnson-Mehl-Kolmogorov (AJMK) model for isothermal conditions and on corresponding modifications such as those of Nakamura [19] and Ozawa [20] for non-isothermal conditions. The authors found in part excellent agreement between these predictions and their own experimental data. For predicting crystallization kinetics under shear conditions the authors used a AJMK model in conjunction with a formulation for the number of activated nuclei under the influence of shear. The latter part of the approach was based on an earlier study of Eder et al. [21,22] who expressed the number of activated nuclei as a function of the square of the shear rate which provides a possibility to predict the thickness of thread-like precursors. A related approach was published by Hieber [23] who used the Nakamura equation to establish a direct relation between the Avrami and the Ozawa crystallization rate constants.

AJMK-based transformation models have been applied to polymers with great success to cases where their underlying assumptions of material homogeneity are reasonably fulfilled [18-23]. It is, however, likely that further progress in understanding and tailoring polymer microstructures can be made by the introduction of automaton models which are designed to cope with more realistic boundary conditions by taking material heterogeneity into account [24]. Important cases where the application of cellular automata is conceivable in the field of polymer crystallization are heterogeneous spherulite formation and growth; heterogeneous topologies and size distributions of spherulites; topological and crystallographic effects arising from additives for heterogeneous nucleation; lateral heterogeneity of the crystallographic texture; overall crystallization kinetics under complicated boundary conditions; reduction of the spherulite size; randomization of the crystallographic texture; as well as effects arising from impurities and temperature gradients.

## 2    Structure of the cellular automaton model and treatment of the crystallographic texture

The model for the mesoscale prediction of spherulite growth in polymers is designed as a 3D cellular automaton model with a probabilistic switching rule. It is discrete in time, real space, and crystallographic orientation space. It is defined on a cubic lattice considering the three nearest neighbor shells for the calculation of cell switches. The kinetics is formulated on the basis of the Hoffman-Lauritzen rate theory for spherulite growth. It is used to calculate the switching probability of each grid point as a function of its previous state and the state of the neighboring grid points. The actual decision about a switching event is made by evaluating the local switching probability using a Monte Carlo step. The growth rule is scaled by the ratio of the local and the maximum possible interface properties, the local and maximum occurring Gibbs free energy of transformation, the local and maximum occurring temperature, and by the spacing of the grid points. The use of experimental input data allows one to make predictions on a real time and space scale. The transformation rule is scalable to any mesh size and to any spectrum of interface and transformation data. The state update of all grid points is made in synchrony.

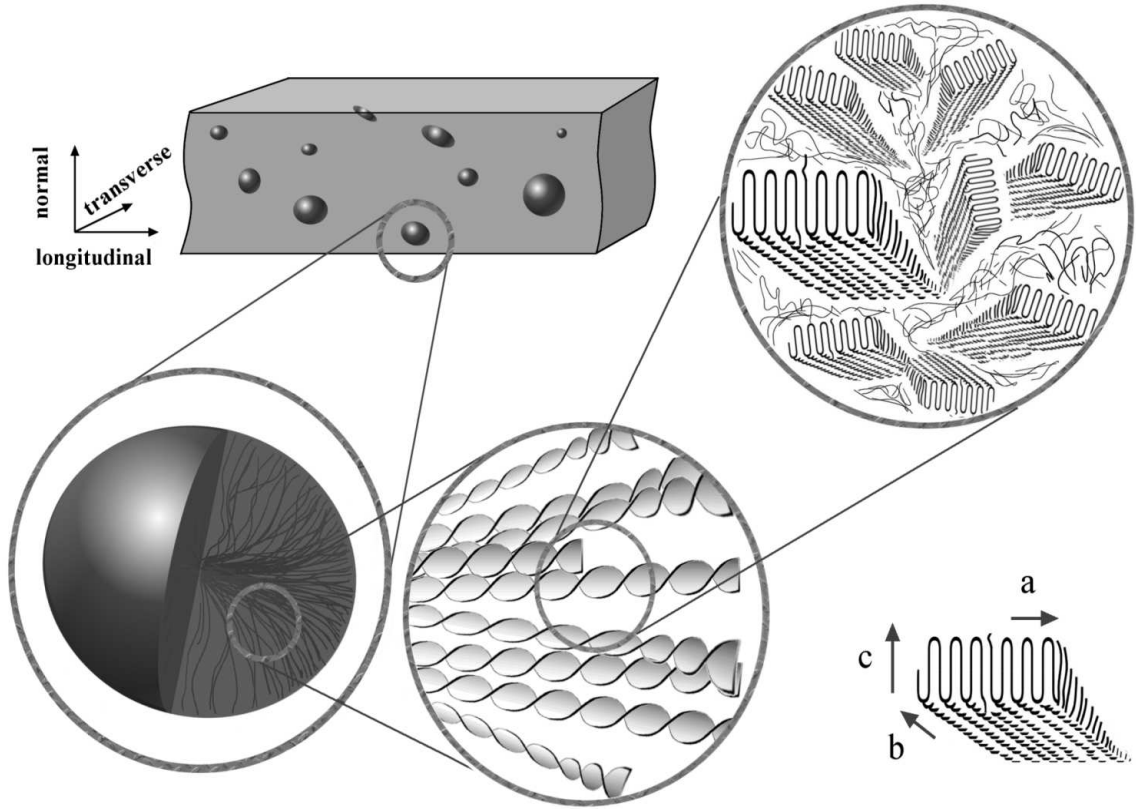

Figure 1

Reference systems for polymer textures (polyethylene as an example). The sample coordinate system is often of orthotropic symmetry as defined by normal, transverse, and longitudinal direction of the macroscopic processing procedure. The reference system of the spherulite can be defined by the crystallographic orientation of the primary nucleus or by the coordinate system inherited by way of epitaxy from an external nucleating agent. Finally, owing to the extensive branching of their internal crystalline lamellae spherulites cannot be characterized by a single orientation but an orientation *distribution*. Further rotational degrees of freedom can add to this complicated crystallographic texture at the single lamella scale due to the molecular conformation of the constituent monomer and the resulting helix angle of the chains (as in the case of polyethylene).

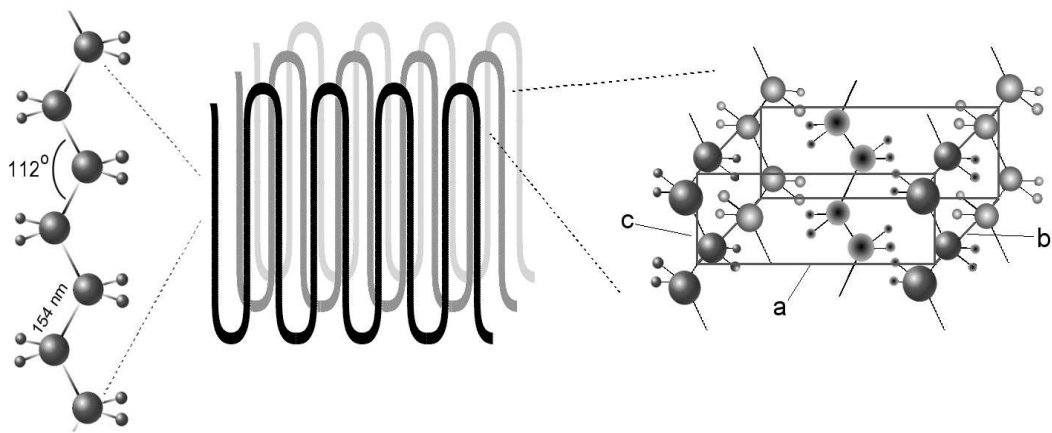

Figure 2

Structure and unit cell of crystalline polyethylene.

Independent variables of the automaton are time $t$ and space $\mathbf{x}=(x_1,x_2,x_3)$. Space is discretized into an array of cubic cells. The state of each cell is characterized in terms of the dependent field variables. These are the temperature field, the phase state (amorphous matrix or partially crystalline spherulite), the transformation and interface energies, the crystallographic orientation distribution of the crystalline nuclei $f(\mathbf{h}) = f(\mathbf{h})(\varphi_1, \phi, \varphi_2)$, and the crystallographic orientation distribution $f(\mathbf{g})=f(\mathbf{g})(\varphi_1, \phi, \varphi_2)$ of one spherulite, where $\mathbf{h}$ and $\mathbf{g}$ are orientation matrices and $\varphi_1, \phi, \varphi_2$ are Euler angles. The introduction of two variable fields for the texture, $\mathbf{h}$ and $\mathbf{g}$, deserves a more detailed explanation: When analyzing crystallographic textures of partially crystalline polymers it is useful to distinguish between different orientational scales. These are defined, first, by the sample coordinate system; second, by the coordinate system of the crystalline nuclei (where the incipient crystal orientation is created by the direction of the alignment of the first lamellae in case of homogeneous nucleation or by epitaxy effects in case of heterogeneous nucleation); third, by the intrinsic helix angle of the single lamellae according to the chain conformation; and fourth, by the complete orientation distribution function of one partially crystalline spherulite (Figures 1, 2). The latter orientation distribution function is characterized by the orientational degrees of freedom provided by the lamella helix angle as well as by orientational branching creating a complete orientation *distribution* in one spherulite which is described by $f(\mathbf{g}) = f(\mathbf{g})(\varphi_1, \phi, \varphi_2)$ relative to its original nucleus of orientation $\mathbf{h}=\mathbf{h}(\varphi_1, \phi, \varphi_2)$. The sample coordinate system is often of orthotropic symmetry as defined by normal, transverse, and longitudinal direction. The reference system of the spherulite can be defined by the crystallographic orientation of the primary nucleus or by the coordinate system inherited by way of epitaxy from an external nucleus in case of heterogeneous nucleation. The orientation of such a nucleus is described by the rotation matrix $\mathbf{h}=\mathbf{h}(\varphi_1, \phi, \varphi_2)$ as a spin relative to the sample coordinate system. The orientation distribution of *all* nuclei amounts to $f(\mathbf{h}) = f(\mathbf{h})(\varphi_1, \phi, \varphi_2)$. When assuming that all spherulites branch in a similar fashion, leading to a single-spherulite texture, $f(\mathbf{g})=f(\mathbf{g})(\varphi_1, \phi, \varphi_2)$, the total orientation distribution of a bulk polymer can then be calculated according to $f(\mathbf{h}) \cdot f(\mathbf{g})$.

The current simulation takes a mesoscopic view at this multi-scale problem, i.e. it does not provide a generic simulation of the texture evolution of a single spherulite but includes only the orientation distribution $f(\mathbf{h})=f(\mathbf{h})(\varphi_1, \phi, \varphi_2)$ of the texture of the nuclei.

The driving force is the Gibbs free energy $G_t$ per switched cell associated with the transformation. The starting data, i.e. the mapping of the amorphous state, the temperature field, and the spatial distribution of the Gibbs free energy, must be provided by experiment or theory.

The kinetics of the automaton result from changes in the state of the cells which are hereafter referred to as cell switches. They occur in accord with a switching rule which determines the individual switching probability of each cell as a function of its previous state and the state of its neighbor cells. The switching rule used in the simulations discussed below is designed for the simulation of static crystallization of a quiescent supercooled amorphous matrix. It reflects that the state of a non−crystallized cell belonging to an amorphous region may change due to the expansion of a crystallizing neighbor spherulite which grows according to the local temperature, Gibbs free energy associated with that transformation, and interface properties. If such an expanding spherulite sweeps a non−crystallized cell the energy of that cell changes and a new orientation distribution is assigned to it, namely that of the growing neighbor spherulite.

## 3 Basic rate equation for spherulitic growth

In 1957 Keller reported that polyethylene formed chainfolded lamellar crystals from solution [25]. This discovery was followed by the confirmation of the generality of this morphology, namely, that lamellar crystals form upon crystallization, both from the solution and from the melt [26] for a wide variety of polymers.

Most theories of polymer crystallization go back to the work of Turnbull and Fisher [27] in which the rate of nucleation is formulated in terms of the product of two Boltzmann expressions, one quantifying the mobility of the growth interface in terms of the activation energy for motion, the other one involving the energy for secondary nucleation on existing crystalline lamellae. On the basis of this pioneering work Hoffman and Lauritzen developed a more detailed rate equation for spherulite growth under consideration of secondary surface nucleation [28]. Overviews on spherulite growth kinetics were given by Keller [29], Hoffman et al. [30], Snyder and Marand [31,32], Hoffman and Miller [33], and Long et al. [34].

The rate equation by Hoffman and Lauritzen is used as a basis for the formulation of the switching rule of the cellular automaton model. It describes growth interface motion in terms of lateral and forward molecule alignment and disalignment processes at a homogeneous planar portion of an interface segment between a semi-crystalline spherulite and the amorphous matrix material. It is important to note that the Hoffman and Lauritzen rate theory is a coarse-grained formulation which homogenizes over two rather independent sets of very anisotropic mechanisms, namely, secondary nucleation on the lateral surface of the growing lamellae (creating strong lateral, i.e. out-of-plane volume expansion and new texture components) and lamella growth (creating essentially 2-dimensional, i.e. in-plane expansion and texture continuation). Although the two basic ingredients of this rate formulation are individually of a highly anisotropic character the overall spherulite equation is typically used in an isotropic form. The basic rate equation is

$$\dot{\mathbf{x}} = \dot{\mathbf{x}}_p \ \exp\left(-\frac{Q^*}{R\left(T - T_\infty\right)}\right) \exp\left(-\frac{K_g}{T \, \Delta T}\right) \tag{1}$$

where $\dot{\mathbf{x}}$ is the velocity vector of the interface between the spherulite and the supercooled amorphous matrix, $\dot{\mathbf{x}}_p$ is the pre-exponential velocity vector, $\Delta T$ the supercooling defined by $\Delta T = T_m^0 - T$, where $T_m^0$ is the equilibrium melting point (valid for a very large crystal formed from fully extended chains, no effect of free surfaces), $Q^*$ is the activation energy for viscous molecule flow or respectively attachment of the chain to the crystalline surface, $T_\infty$ is the temperature below which all viscous flow stops (glass transition temperature, temperature at which the viscosity exceeds the value of $10^{13}$ Ns/m$^2$), $K_g$ the secondary nucleation exponent, and $T$ the absolute temperature of the (computer) experiment.

According to Hoffman and Lauritzen [28,30,33] the exponent $K_g$ amounts to

$$K_g = \frac{\xi \, b \, \sigma \, \sigma_e \, T_m^0}{k_B \Delta G_f} \tag{2}$$

where $k_B$ is the Boltzmann constant, $\xi$ a constant which equals 4 for growth regimes I and III and 2 for growth regime II [33,35], $b$ is the thickness of a crystalline lattice cell in growth direction (see Figures 1,2), $\sigma$ is the free energy per area for the interface between the lateral surface and the supercooled melt, $\sigma_e$ is the free energy per area for the interface between the fold surface where the molecule chains fold back or emerge from the lamella and the supercooled melt, and $\Delta G_f$ is the Gibbs free energy of fusion at the crystallization temperature which is approximated by $\Delta G_f \approx \left( \Delta H_f \, \Delta T \right) / T_m$ [28,30,33].

The different growth regimes, I (shallow quench regime), II (deep quench regime), and III (very deep quench regime), were discussed by Hoffman [33] in the sense that within regime I of spherulite growth the expansion of the spherulite is dominated by weak secondary nucleation and preferred forward lamellae growth. Nucleation is in this regime dominated by the formation of few single surface nuclei which are referred to as stems. It is assumed that when one stem is nucleated the entire new layer is almost instantaneously completed relative to the nucleation rate. In regime II the two rates are comparable. Regime II is the normal condition for spherulitic growth. It has a weaker $\Delta T$ dependence than regime I. In regime III nucleation of many nuclei occurs leading to disordered crystal growth. Nucleation is in this regime more important than growth.

According to the work of Hoffman and Miller [33] the pre-exponential velocity vector for an interface between a spherulite and the supercooled amorphous matrix can be modeled according to

$$\dot{\mathbf{x}}_p = \mathbf{n} \, \frac{N_0 \, b \, J}{\ell_u} \left[ \frac{k_B T}{b \, \sigma} - \frac{k_B T}{2 \, b \, \sigma + a \, b \, \Delta G_f} \right] \qquad (3)$$

where $N_0$ is the number of initial stems, soon to be involved in the first stem deposition process, $\mathbf{n}$ the unit normal vector of the respective interface segment, and $\ell_u$ the monomer length. $J$ amounts to

$$J = \frac{\varsigma}{n_z} \left( \frac{k_B T}{h} \right) \qquad (4)$$

where $\varsigma$ is a constant for the molecular friction experienced by the chain as it is reeled onto the growth interface, $n_z$ the number of molecular repeat units in the chain, and $h$ the Planck constant. The fraction $k_B T / h$ can be interpreted as a frequency pre-factor. The pre-exponential velocity vector can then be written

$$\dot{\mathbf{x}}_p = \mathbf{n} \, \frac{N_0 \, b \, \varsigma}{\ell_u \, n_z} \left( \frac{k_B T}{h} \right) \left[ \frac{k_B T}{b \, \sigma} - \frac{k_B T}{2 \, b \, \sigma + a \, b \, \Delta G_f} \right] \qquad (5)$$

yielding a Hoffman-Lauritzen-Miller type version [33] of a velocity-rate vector equation for spherulite growth under consideration of secondary nucleation

$$\dot{\mathbf{x}} = \mathbf{n}\,\frac{N_0\,b\,\varsigma}{\ell_u\,n_z}\left(\frac{k_\mathrm{B}T}{h}\right)\left[\frac{k_\mathrm{B}T}{b\,\sigma} - \frac{k_\mathrm{B}T}{2\,b\,\sigma + a\,b\,\Delta G_f}\right]\exp\left(-\frac{Q^*}{R\left(T - T_\infty\right)}\right)\cdot$$
$$\cdot\exp\left(-\frac{\xi\,b\,\sigma\,\sigma_\mathrm{e}\,T_m^0}{k_\mathrm{B}\,T\,\Delta T\,\Delta G_f}\right) \tag{6}$$

# 4   Using rate theory as a basis for the automaton model

## 4.1   Mapping the rate equation on a cellular automaton mesh

For dealing with competing state changes affecting the same lattice cell in a cellular automaton, the statistical rate equation can be rendered into a probabilistic analogue which allows one to calculate switching probabilities for the states of the cellular automaton cells [2]. For this purpose, the rate equation is separated into a non-Boltzmann part, $\dot{\mathbf{x}}_0$, which depends comparatively weakly on temperature, and a Boltzmann part, $w$, which depends exponentially on temperature.

$$\dot{\mathbf{x}} = \dot{\mathbf{x}}_0\,w = \mathbf{n}\,\frac{N_0\,b\,\varsigma}{\ell_u\,n_z}\left(\frac{k_\mathrm{B}T}{h}\right)\left[\frac{k_\mathrm{B}T}{b\,\sigma} - \frac{k_\mathrm{B}T}{2b\,\sigma + ab\Delta G_f}\right]\exp\left(-\frac{Q^*}{R\left(T - T_\infty\right)}\right)\cdot$$
$$\cdot\exp\left(-\frac{\xi\,b\,\sigma\,\sigma_\mathrm{e}\,T_m^0}{k_\mathrm{B}\,T\,\Delta T\,\Delta G_\mathrm{f}}\right)$$

$$(7)$$

$$\text{with} \quad \dot{\mathbf{x}}_0 = \mathbf{n}\,\frac{N_0\,b\,\varsigma}{\ell_u\,n_z}\left(\frac{k_\mathrm{B}T}{h}\right)\left[\frac{k_\mathrm{B}T}{b\,\sigma} - \frac{k_\mathrm{B}T}{2\,b\,\sigma + a\,b\,\Delta G_f}\right]$$

$$\text{and} \quad w = \exp\left(-\frac{Q^*}{R\left(T - T_\infty\right)}\right)\exp\left(-\frac{\xi\,b\,\sigma\,\sigma_\mathrm{e}\,T_m^0}{k_\mathrm{B}\,T\,\Delta T\,\Delta G_\mathrm{f}}\right)$$

The Boltzmann factors, $w$, represent the probability for cell switches. According to this equation non–vanishing switching probabilities occur for cells with different temperatures and/or transformation energies. The automaton considers the first, second (2D), and third (3D) neighbor shell for the calculation of the switching probability acting on a cell. The local value of the switching probability may in principal depend on the crystallographic character of the interfaces entering via the interface energies in the secondary nucleation term. The current simulation does, however, not account for this potential source of anisotropy but assumes constant values for the interface energies of the lateral and of the fold interface.

## 4.2    The scaled and normalized switching probability

The cellular automaton is usually applied to starting data which have a spatial resolution far above the atomic scale. This means that the automaton grid may have some mesh size $\lambda_m >> b$. If a moving boundary segment sweeps a cell, the spherulite thus grows (or shrinks) by $\lambda_m^3$ rather than $b^3$. Since the net velocity of an interface segment (between a partially crystalline spherulite and the amorphous matrix) must be independent of the imposed value of $\lambda_m$, an increase of the jump width must lead to a corresponding decrease of the grid attack frequency, i.e. to an increase of the characteristic time step, and vice versa. For obtaining a scale–independent interface velocity, the grid frequency must be chosen in a way to ensure that the attempted switch of a cell of length $\lambda_m$ occurs with a frequency much below the atomic attack frequency which attempts to switch a cell of length b. Mapping the rate equation on such an imposed grid which is characterized by an external scaling length (lattice parameter of the mesh) $\lambda_m$ leads to the equation

$$\dot{\mathbf{x}} = \dot{\mathbf{x}}_0 w = \mathbf{n}(\lambda_m \nu) w \quad \text{with} \quad \nu = \frac{N_0\, b\, \varsigma}{\lambda_m\, \ell_u\, n_z}\left(\frac{k_B T}{h}\right)\left[\frac{k_B T}{b\,\sigma} - \frac{k_B T}{2\, b\,\sigma + a\, b\, \Delta G_f}\right] \qquad (8)$$

where $\nu$ is the eigenfrequency of the chosen mesh characterized by the scaling length $\lambda_m$.

The eigenfrequency given by this equation represents the attack frequency which is valid for *one* particular interface with constant self-reproducing properties moving in a constant temperature field. In order to introduce the possibility of a whole *spectrum* of interface properties (due to temperature fields and intrinsic interface energy anisotropy) in one simulation it is necessary to normalize this equation by a general grid attack frequency $\nu_0$ which is *common* to all interfaces in the system rendering the above equation into

$$\dot{\mathbf{x}} = \dot{\mathbf{x}}_0\, w = \mathbf{n}\, \lambda_m\, \nu_0\left(\frac{\nu}{\nu_0}\right) w = \hat{\dot{\mathbf{x}}}_0\left(\frac{\nu}{\nu_0}\right) w = \hat{\dot{\mathbf{x}}}_0\, \hat{w} \qquad (9)$$

where the normalized switching probability amounts to

$$
\begin{aligned}
\hat{w} &= \left(\frac{\nu}{\nu_0}\right)\exp\left(-\frac{Q^*}{R(T-T_\infty)}\right)\exp\left(-\frac{\xi\,b\,\sigma\,\sigma_e\,T_m^0}{k_B\,T\,\Delta T\,\Delta G_f}\right) \\
&= \frac{N_0\,b\,\varsigma}{\nu_0\,\lambda_m\,\ell_u\,n_z}\left(\frac{k_B T}{h}\right)\left[\frac{k_B T}{b\,\sigma} - \frac{k_B T}{2\,b\,\sigma + a\,b\,\Delta G_f}\right]\exp\left(-\frac{Q^*}{R(T-T_\infty)}\right)\cdot \\
&\qquad\qquad\qquad\qquad\qquad\qquad \cdot\exp\left(-\frac{\xi\,b\,\sigma\,\sigma_e\,T_m^0}{k_B\,T\,\Delta T\,\Delta G_f}\right)
\end{aligned}
\tag{10}
$$

The value of the normalization or grid attack frequency $\nu_0$ can be identified by using the plausible assumption that the maximum occurring switching probability can not assume a state above one

$$
\begin{aligned}
\hat{w}^{max} &= \frac{N_0\,b\,\varsigma}{\nu_0\,\lambda_m\,\ell_u\,n_z}\left(\frac{k_B T_{max}}{h}\right)\left[\frac{k_B T_{max}}{b\,\sigma_{min}} - \frac{k_B T_{max}}{2\,b\,\sigma_{min} + a\,b\,\Delta G_f}\right]\cdot \\
&\qquad \cdot\exp\left(-\frac{Q^*}{R(T_{max}-T_\infty)}\right)\exp\left(-\frac{\xi\,b\,\sigma_{min}\,\sigma_{e,min}\,T_m^0}{k_B\,T_{max}\,\Delta T_{max}\,\Delta G_{f,max}}\right) \overset{!}{\leq} 1
\end{aligned}
\tag{11}
$$

where $T_{max}$ is the maximum occurring temperature in the system, $\Delta T_{max}$ the maximum occurring supercooling, $\sigma_{min}$ the minimum occurring lateral interface energy (for instance in cases where a crystallographic orientation dependence of the interface energy exists), $\Delta G_{f,max}$ the maximum Gibbs free transformation energy which depends on the local temperature, i.e. $\Delta G_{f,max} \approx (\Delta H\,\Delta T_{max})/T_m^0$, and $\sigma_{e,min}$ the minimum occurring fold interface energy.

With $\hat{w}^{max} = 1$ in the above equation one obtains the normalization frequency, $\nu_0^{min}$, as a function of the upper bound input data.

$$
\begin{aligned}
\nu_0^{min} &= \frac{N_0\,b\,\varsigma}{\lambda_m\,\ell_u\,n_z}\left(\frac{k_B T_{max}}{h}\right)\left[\frac{k_B T_{max}}{b\,\sigma_{min}} - \frac{k_B T_{max}}{2b\,\sigma_{min} + ab\Delta G_f}\right]\cdot \\
&\qquad \cdot\exp\left(-\frac{Q^*}{R(T_{max}-T_\infty)}\right)\exp\left(-\frac{\xi\,b\,\sigma_{min}\,\sigma_{e,min}\,T_m^0}{k_B\,T_{max}\,\Delta T_{max}\,\Delta G_{f,max}}\right)
\end{aligned}
\tag{12}
$$

This frequency must only be calculated once per simulation. It normalizes all other switching processes.

Inserting this basic attack frequency of the grid into equation (10) yields an expression for calculating the local values of the switching probability as a function of temperature and energy

$$
\hat{w}^{\text{local}} = \left(\frac{T_{\text{local}}}{T_{\text{max}}}\right) \frac{\left[\dfrac{k_{\text{B}} T_{\text{local}}}{b\,\sigma_{\text{local}}} - \dfrac{k_{\text{B}} T_{\text{local}}}{2\,b\,\sigma_{\text{local}} + a\,b\,\Delta G_{\text{f,local}}}\right]}{\left[\dfrac{k_{\text{B}} T_{\text{max}}}{b\,\sigma_{\text{min}}} - \dfrac{k_{\text{B}} T_{\text{max}}}{2\,b\,\sigma_{\text{min}} + a\,b\,\Delta G_{\text{f,max}}}\right]} \cdot
$$
$$
\cdot \exp\left(-\left[\left(\frac{Q^{*}}{R}\right)\left(\frac{1}{(T_{\text{local}} - T_{\infty})} - \frac{1}{(T_{\text{max}} - T_{\infty})}\right)\right]\right) \cdot \tag{13}
$$
$$
\cdot \exp\left(-\left[\left(\frac{\xi\,b\,T_m^0}{k_{\text{B}}}\right)\left(\frac{\sigma_{\text{local}}\,\sigma_{\text{e,local}}}{T_{\text{local}}\,\Delta T_{\text{local}}\Delta G_{\text{f,local}}} - \frac{\sigma_{\text{min}}\,\sigma_{\text{e,min}}}{T_{\text{max}}\,\Delta T_{\text{max}}\Delta G_{\text{f,max}}}\right)\right]\right)
$$

This expression is the central switching equation of the algorithm. The cellular automaton generally works in such a way that an existing (expanding) spherulite *infects* its neighbor cells at a rate or respectively probability which is determined by equation (13). This means that wherever a partially crystalline spherulite has the topological possibility to expand into an amorphous neighbor cell the automaton rule uses equation (13) to quantify the probability of that cell switch using the local state variable data which characterize the two neighboring cells.

Equation (13) reveals that local switching probabilities on the basis of rate theory can be quantified in terms of the ratio of the local and the maximum temperature and the corresponding ratio of the interface properties. The probability of the fastest occurring interface segment to realize a cell switch is exactly equal to 1. The above equation also shows that the mesh size of the automaton does not influence the switching probability but only the time step elapsing during an attempted switch. The characteristic time constant of the simulation $\Delta t$ is $1/\nu_0^{\text{min}}$.

In this context the local switching probability can also be regarded as the ratio of the distances that can be swept by the local interface and the interface with maximum velocity, or as the number of time steps the local interface needs to wait before crossing the encountered neighbor cell [6]. This reformulates the same underlying problem, namely, that interfaces with different velocities cannot switch the state of the automaton in a given common time step with the same probability.

In the current automaton formulation competing cell switches aiming at transforming the same cell state are each considered by a stochastic decision rather than a counter to account for insufficient sweep of the boundary through the cell. Stochastic Markov–type sampling is equivalent to installing a counter, since the probability to switch the automaton is proportional to the velocity ratio given by the above equations, provided the chosen random number generator is truly stochastic.

### 4.3 The switching decision

According to the argumentation given above equation (13) is used to calculate the switching *probability* of a cell as a function of its previous state and the state of the neighbor cells. The actual *decision* about a switching event for each cell is made by a Monte Carlo step. The use of random sampling ensures that all cells are switched according to their proper statistical weight, i.e. according to the local driving force and mobility between cells. The simulation proceeds by calculating the individual local switching probabilities $\hat{w}^{local}$ according to equation (13) and evaluating them using a non–Metropolis Monte Carlo algorithm. This means that for each cell the calculated switching probability is compared to a randomly generated number $r$ which lies between 0 and 1. The switch is accepted if the random number is equal or smaller than the calculated switching probability. Otherwise the switch is rejected.

$$\text{random number } r \text{ between 0 and 1} \quad \left\{ \begin{array}{lll} \text{accept switch} & \text{if} & r \leq \hat{w}^{local} \\ \text{reject switch} & \text{if} & r > \hat{w}^{local} \end{array} \right. \quad (14)$$

Except for the probabilistic evaluation of the analytically calculated transformation probabilities, the approach is entirely deterministic. Artificial thermal fluctuation terms other than principally included through the Boltzmann factors are not permitted. The use of realistic or even experimental input data for the interface energies and transformation enthalpies enables one to introduce scale. The switching rule is scalable to any mesh size and to any spectrum of interface energy and Gibbs free transformation energy data. The state update of all cells is made in synchrony, like in all automata [1].

## 4.4    Scaling procedure on the basis of data for polyethylene

Real time and length scaling of the system enters through the physical parameters which characterize the polymer and through the mesh size of the automaton. The central scaling expression is given by equation (12) in conjunction with the terms given by equations (2)-(5). The inverse of the frequency given by equation (12) is the basic time step of the simulation during which each cell has one Monte Carlo attempt to perform a switch in accord with its individual local switching probability as expressed by equation (13). The current simulations use material data for polyethylene taken from [31,32,33], i.e. $T_m^0 = 418.7\,\mathrm{K}$, $Q^* = 5736\,\mathrm{cal/mol}$, $\Delta H_f = 2.8\cdot10^8\,\mathrm{J/m^3}$, $\sigma = 1.18\cdot10^{-2}\,\mathrm{J/m^2}$, $\sigma_e = 9.0\cdot10^{-2}\,\mathrm{J/m^2}$, $a=4.55\,\text{Å}$, $b=4.15\,\text{Å}$, $\ell_u = 1.27\cdot10^{-10}\,\mathrm{m}$, $n_z = 2000$, $T_\infty = 220.0\,\mathrm{K}$, $\varsigma = 2.2\cdot10^{-12}\,\mathrm{kg\cdot s}$, as well as $\lambda_m = 1\,\mu m$ as length of one cubic cellular automaton cell. These data give typical maximum growth velocities of the spherulite at the peak temperature between weak secondary nucleation at large crystallization temperatures and weak diffusion at low crystallization temperatures of the order of $10^{-7} - 10^{-6}$ m/s.

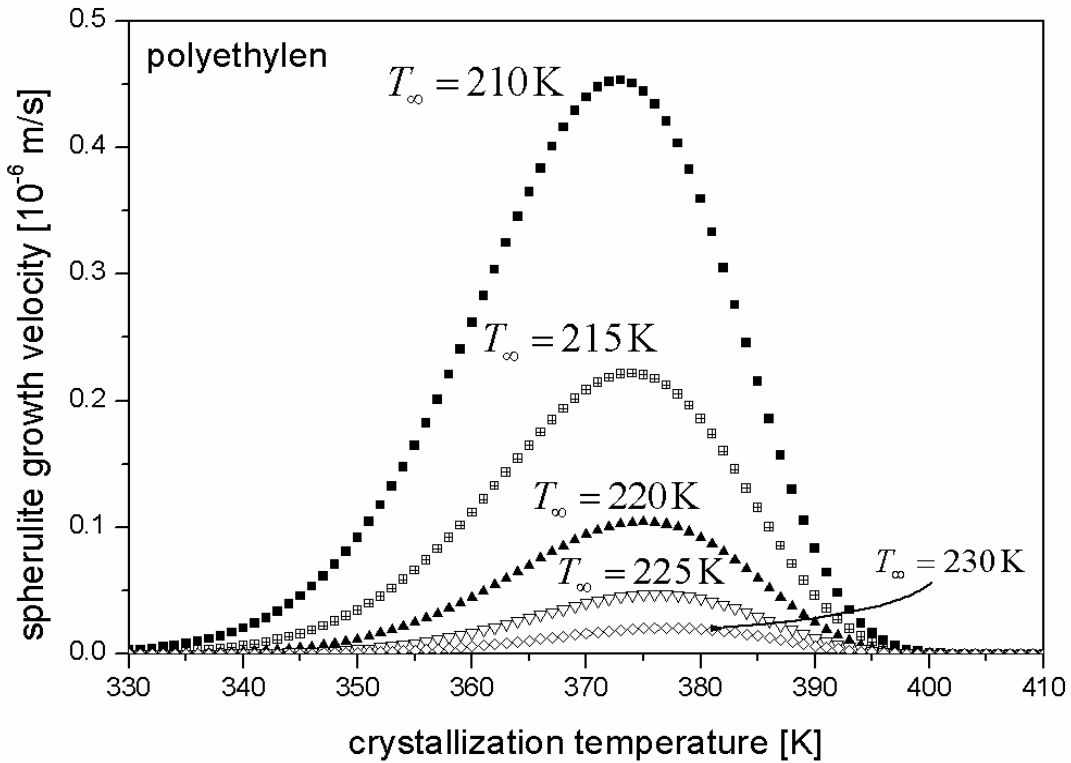

Figure 3
Spherulite growth velocities (polyethylene) for a set of simulations with different assumed values for $T_\infty$.

Polyethylene is a macromolecular solid in which the molecular units are long chain-like molecules with C atoms forming a zig-zag arrangement along the backbone and 2 H atoms attached to each C. Figure 2 gives a schematical presentation of the molecule and the unit cell. The monomer unit of polyethylene is $C_2H_4$. The crystals form by folding the chains alternately up and down and arranging the straight segments between folds into a periodic array. The crystal has orthorhombic body centered symmetry ($a \neq b \neq c$, all angles 90°). Figure 3 shows the growth velocities for a set of simulations with different assumed temperatures for $T_\infty$.

Possible variations in the glass transition temperature are of substantial interest for the simulation of spherulite growth microstructures under complex boundary conditions. The reason for this is that the glass transition temperature can be substantially altered by changing certain structural features or external process parameters. Examples for structural aspects are the dependence of the glass transition temperature on chain flexibility, stiffness, molecular weight, branching, or crystallinity. Examples for external process aspects, which in part drastically influence the aforementioned structural state of the material, are accumulated elastic-plastic deformation, externally imposed shear rates, or the hydrostatic pressure.

## 4.5    Nucleation criteria

Two phenomenological approaches were used in the simulations to treat nucleation, namely site saturated heterogeneous nucleation and constant homogeneous nucleation. The calculations with site saturated nucleation condition were based on the assumption of external heterogeneous nucleation sites such as provided at t=0 s by small impurities or nucleating agent particles. The calculations with constant thermal nucleation were conducted using the nucleation rate equation of Hoffman et al. [28,30].

# 5 Simulation results and discussion

## 5.1 Kinetics and spherulite structure for site saturated conditions

This section presents some 3D simulation results for spherulite growth under the assumption of site saturated nucleation conditions using a cellular automaton with 10 million lattice cells. These standard conditions are chosen to study the reliability of the new method with respect to the kinetic exponents (comparison with the analytical Avrami-Johnson-Mehl-Kolmogorov solution), to lattice effects (spatial discreteness), to topology, and to statistics.

Figures 4 and 5 show the kinetics for simulations at 375 K and 360 K, respectively. The calculations were conducted with 0.05%, 0.01%, 0.005%, and 0.001% of all cells as initial nucleation sites (site saturated). The figures show the volume fractions occupied by spherulites as a function of the isothermal heat treatment time. The resulting topology of the spherulites is given in Figure 6. It is important to note in this context that the depicted spherulite volume fraction must not be confused with the *crystalline* volume fraction, since the spherulites are two-phase aggregates consisting of heavily branched crystalline lamellae with amorphous chains between them (Figure 1). The kinetic anisotropy which is principally inherent in such structures at the nanoscopic scale in terms of secondary nucleation events on the lateral lamellae surfaces and of in-plane lamella growth [36,37] is homogenized at the mesoscopic scale, where the growing spherulites typically behave as isotropic spheres, equation (1). In the present study this results in Avrami-type kinetics with an exponent of 3.0±0.1%. The small observed deviations of ±0.1% from the ideal kinetic exponent of 3.0 occur at small and at large times due to the influence of the discreteness of the cubic automaton lattice at these early and late growth stages, respectively. The scatter in the data for the spherulite volumes at large times (particularly above 99% spherulite volume) is even more important than that for short times (see Figures 4b, 5b). This effect is essentially due to the fact that the transformation from the amorphous to the spherulite state performed by the last cell switches becomes increasingly discrete owing to the changing ratio between the final non-transformed volume and the switched cells. The relevance of the data predicted for the final stages of the transformation is, therefore, somewhat overemphazised and should be treated with some skepticism for instance when analyzing kinetical exponents at the end of the transformation.

It should also be mentioned that three simulation runs were conducted for each set of starting conditions in order to inspect statistical effects arising from the Monte Carlo integration scheme, see section 4.3 and equation (14). Figures 4 and 5 reveal that the statistical fluctuations arising from this part of the simulation procedure are very small (note the similarity of three curves on top of each other in both sets of figures).

The simulated spherulite growth kinetics are in good qualitative accord with experimental data from the literature [38-40]. It is important to underline in this context though that the comparison of the experimentally observed literature data [38-40] with the here simulated crystallization kinetics remains at this stage rather imprecise. This is due to the fact that some of the simulation parameters as required in the current simulations such as for instance the nucleation rates, the cooling rates, and the exact structural data of the used PE specimens were not known or not sufficiently documented in the corresponding publications, at least not in the depth required for this type of simulation. In a second step it is, however, absolutely conceivable to conduct real one-to-one comparisons between simulation and experiment since some recent publications provide more detailed experimental data [e.g. see 41,42].

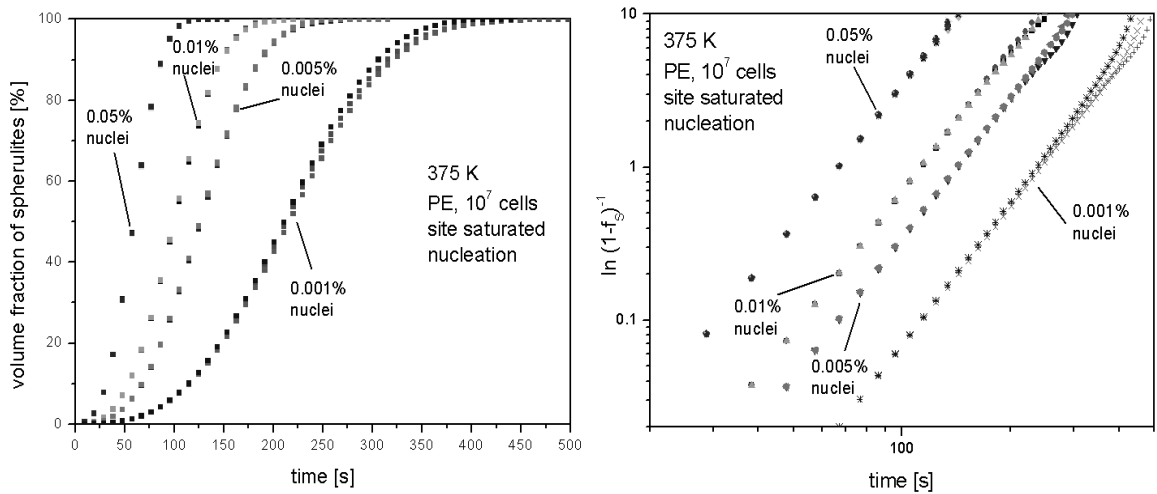

Figure 4
Avrami analysis of the volume fraction occupied by spherulites ($f_s$) as a function of time; site saturated nucleation conditions; polyethylene; 375 K; $10^7$ cells.

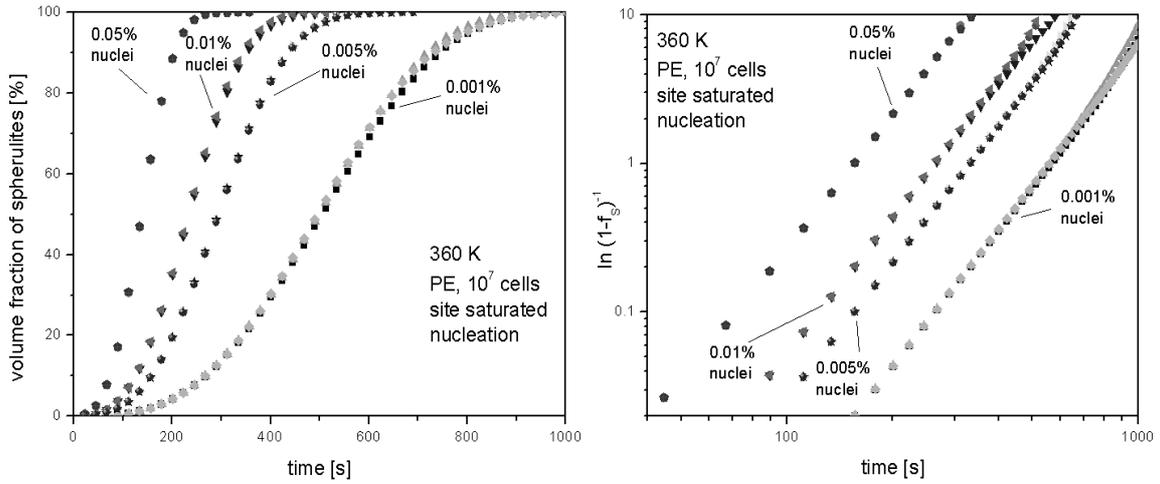

Figure 5
Avrami analysis of the volume fraction occupied by spherulites ($f_s$) as a function of time; site saturated nucleation conditions; polyethylene; 360 K; $10^7$ cells.

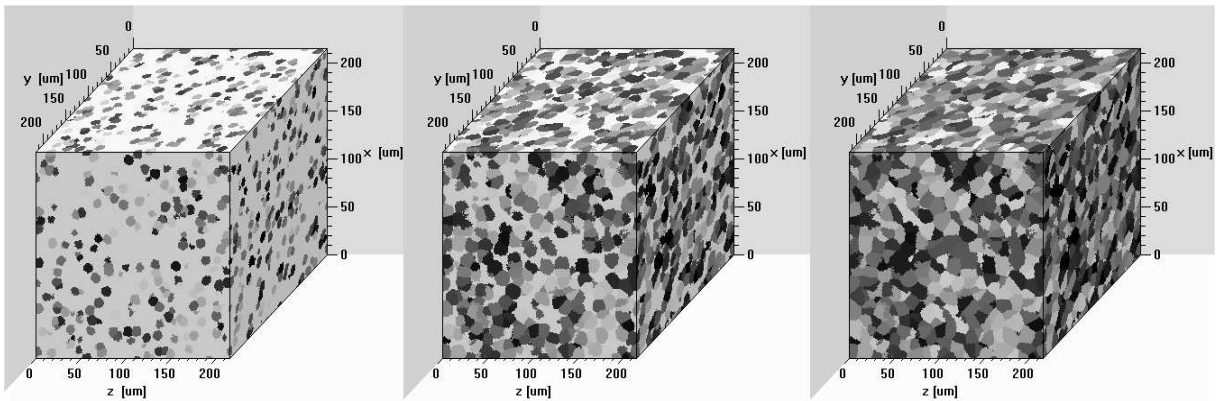

Figure 6
Three subsequent sketches of the simulated spherulite microstructure at 375 K for $10^3$ site saturated nucleation sites (0.05% of $10^7$ cells). The gray scale indicates the respective rotation matrices $\mathbf{h}=\mathbf{h}(\varphi_1, \phi, \varphi_2)$ of the spherulite *nuclei* (not of the entire spherulites) expressed in terms of their rotation angle relative to the sample reference system neglecting the rotation axis. The residual volume (white) is amorphous.

Figure 7 shows the corresponding Cahn-Hagel diagrams for the 2 simulations presented in Figures 4 and 5. Cahn-Hagel diagrams quantify the ratio of the total interfacial area of all spherulites with the residual amorphous matrix and the sample volume as a function of the spherulitic volume fraction. For an analytical Avrami-type case and site saturated conditions this curve assumes a maximum at 50% spherulite growth which is well fulfilled for the present simulations.

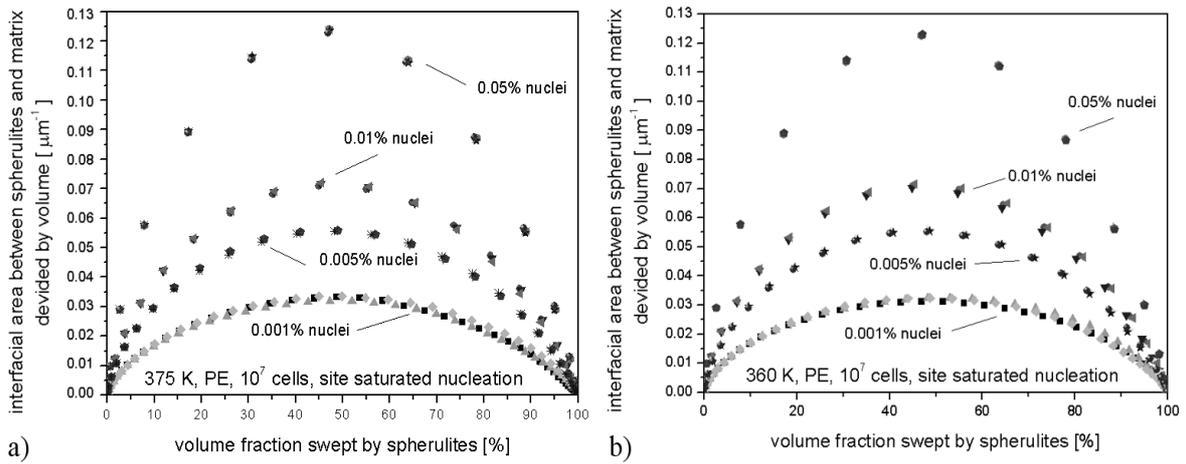

Figure 7
Cahn-Hagel diagram; total interfacial area between the spherulitic material and the residual amorphous matrix divided by the sample volume as a function of the spherulitic volume fraction; site saturated nucleation conditions; polyethylene; $10^7$ cells.
a) 375 K; b) 360 K

Figure 8 shows the resulting spherulite size distributions in terms of the spherulite volumes for the four cases 0.001% cells as initial nuclei (a), 0.005% cells as initial nuclei (b), 0.01% cells as initial nuclei (c), and 0.05% cells as initial nuclei (d) (site saturated conditions). The diagrams use a logarithmic axis for the spherulite size classes and a normalized axis for the spherulite frequencies (number of spherulites in each size class divided by the *total* number of spherulites). Such a presentation provides a good comparability among the four simulation sets. The results were fitted by using a logarithmic normal distribution (solid line in each diagram) which is usually fulfilled for Avrami-type growth processes with site-saturated nucleation conditions. The comparison shows that the simulations indeed reproduce the statistical topological behavior of Avrami processes very well.

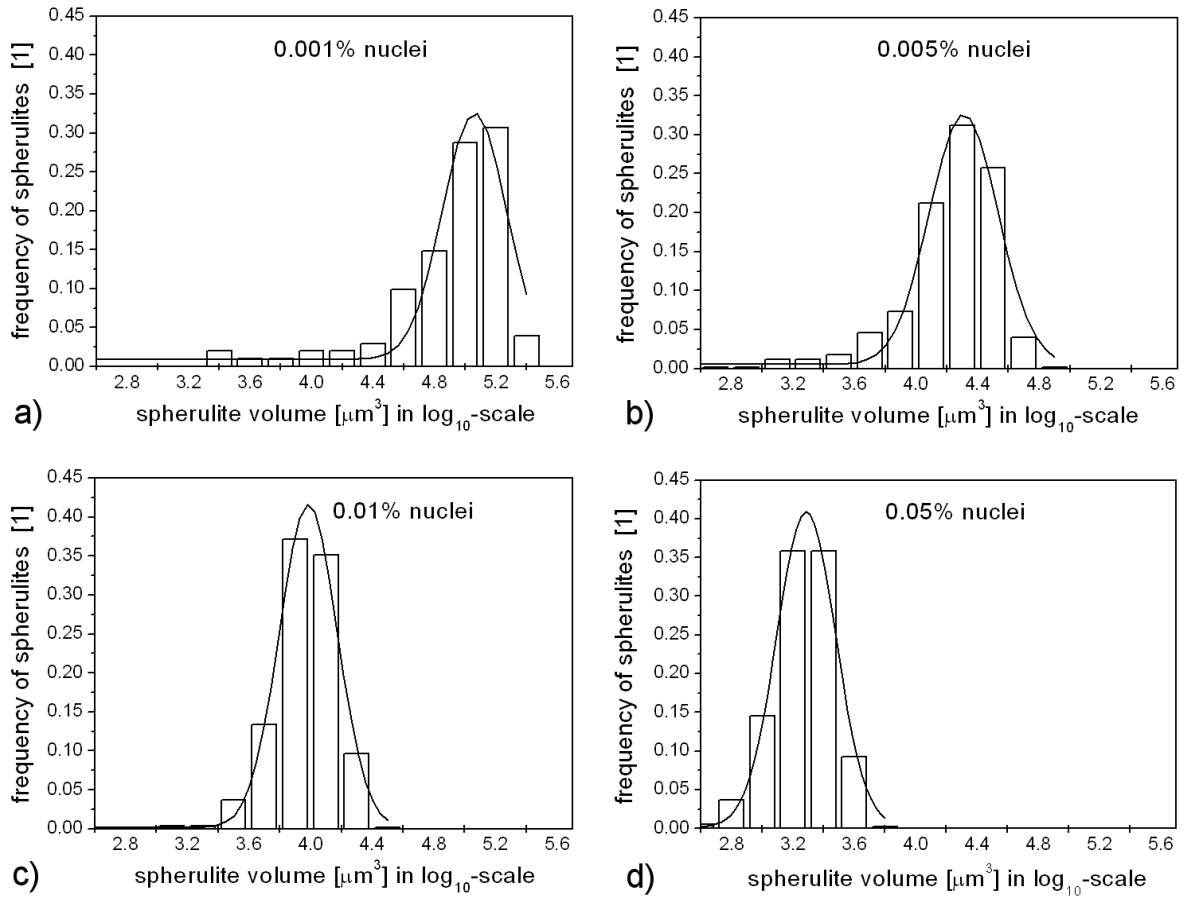

Figure 8
Spherulite size distributions in terms of the spherulite volumes for four different site-saturated nucleation conditions using a logarithmic axis for the spherulite size classes and a normalized axis for the spherulite frequencies (number of spherulites in each size class divided by the *total* number of spherulites). The lines represent curve fits by use of a logarithmic normal distribution.
a) 0.001% nuclei; b) 0.005% nuclei; c) 0.01% nuclei; d) 0.05% nuclei.

The data nicely document the gradual shift of the final spherulite size from conditions with a very small number of initial nuclei (Figure 8a, 0.001% cells as initial nuclei, large average spherulite size after heat treatment) to conditions with a very large number of initial nuclei (Figure 8d, 0.05% cells as initial nuclei, smaller average spherulite size after heat treatment).

## 5.2 Kinetics and spherulite structure for constant nucleation rate

The simulations were also conducted for constant nucleation rate using a set of different activation energies for nucleation. In the present study this results in Avrami-type kinetics with an exponent of 4.0±1%. This deviation of ±1% from the ideal analytical exponent of 4 represents a rather large scatter which, however, can be attributed to the discreteness of the cubic automaton lattice. The fact that the deviation in kinetics (±1%) is much larger than that observed for the simulations with site saturated nucleation conditions (±0.1%) (section 5.1, Figures 4,5) can be explained by the temporal change in the ratio between the remaining amorphous matrix material which is not yet swept and the new nucleation cells. This means that – since the residual matrix volume which is not swept is becoming smaller with each simulation step – each new nucleus which is added to the lattice during one time step occupies an increasingly larger finite volume relative to the rest of the material. The analytical result of 4, however, is based on the assumption of a vanishing volume of new nucleation sites. Figure 10a shows three subsequent microstructures of the same simulation. Figure 10b shows results form a corresponding set of simulation results with site saturated conditions, but under the influence of a temperature gradient (340 K – 375 K) along the y-direction.

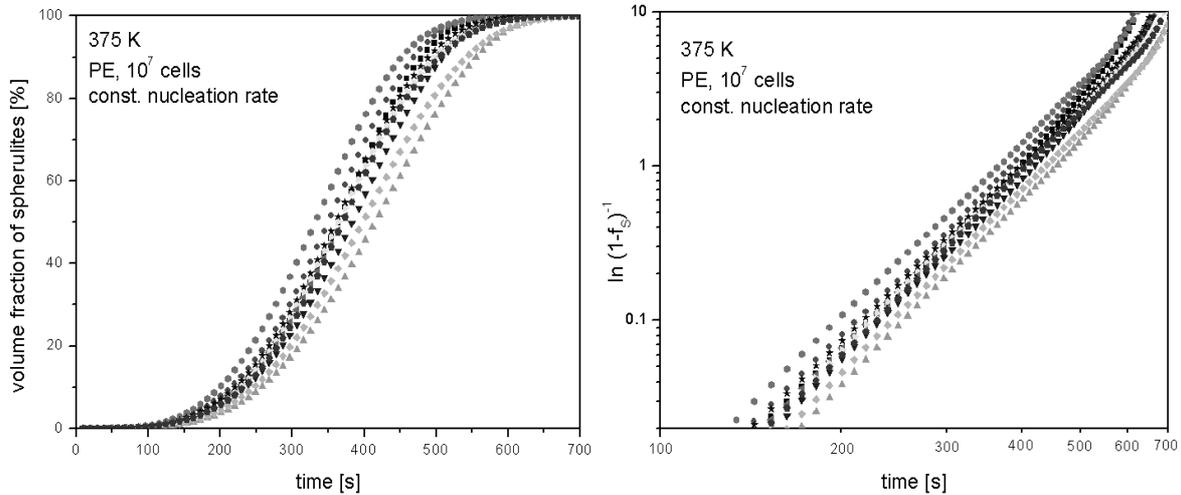

Figure 9
Volume fraction occupied by spherulites for simulations with constant nucleation rate (different nucleation energies) as a function of time, polyethylene, 375 K, automaton with $10^7$ cells.

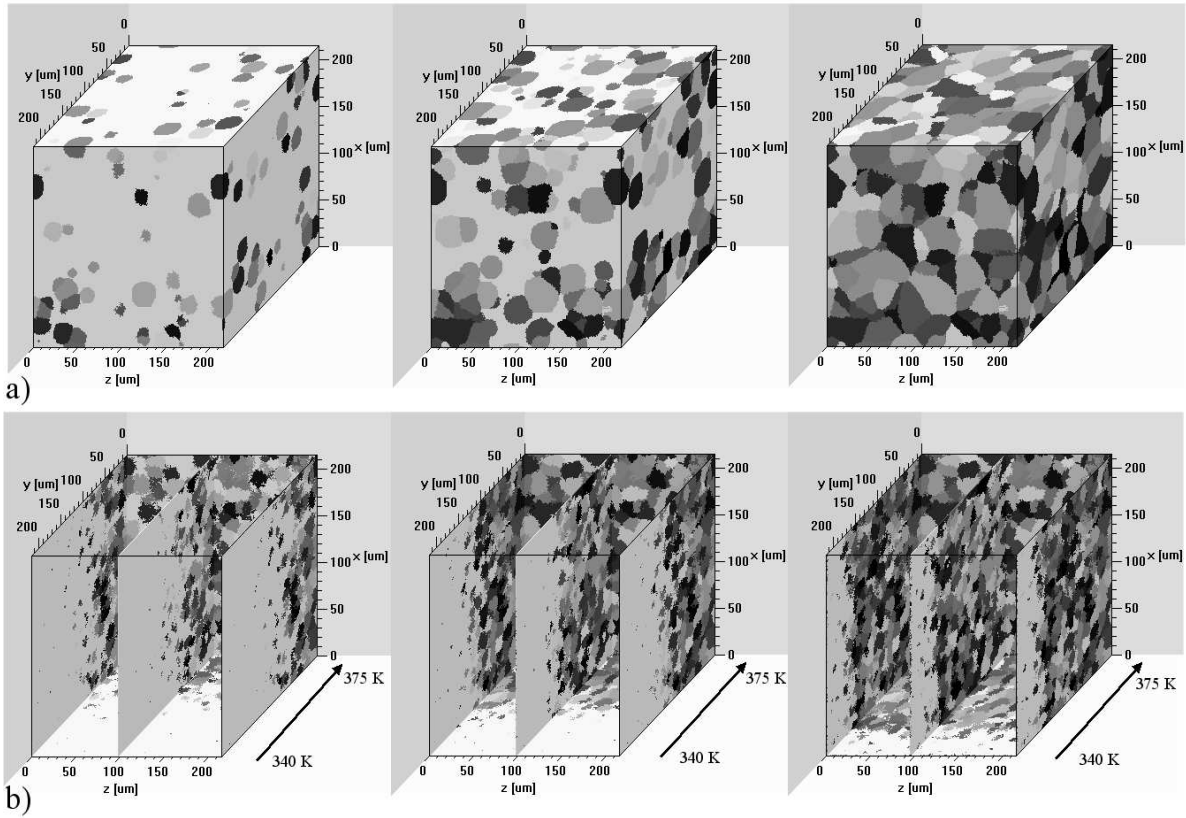

Figure 10
Subsequent sketches of simulated spherulite microstructures for constant nucleation rate. The gray scale follows the crystalline orientations of the spherulite *nuclei* as in Figure 6. The residual volume (white) is amorphous.
a) 375 K  b) influence of a temperature gradient (340 K – 375 K) along the y-direction.

# Conclusions

A new 3D cellular automaton model with a Monte Carlo switching rule was introduced to simulate spherulite growth in polymers using polyethylene as a model substance [46,47]. The automaton is discrete in time, real space, and orientation space. The switching probability of the lattice points is formulated according to the kinetic theory of Hoffman and Lauritzen. It is scaled by the ratio of the local and the maximum interface energies, the local and maximum occurring Gibbs free energy of transformation, the local and maximum occurring temperature, and by the spacing of the lattice points. The use of experimental input data for polyethylene provides a real time and space scale. The article shows that the model is capable of correctly reproducing 3D Avrami-type kinetics for site saturated nucleation and constant nucleation rate under isothermal and temperature-gradient conditions. The simulated kinetic results showed good correspondence to experimental data from the literature. Corresponding predictions of Cahn-Hagel diagrams also matched analytical models. Besides these basic verification exercises the study aimed to communicate that the main advantage of the new model is that it can offer more details than Avrami-type approximations which are typically used in this field. In particular, the new cellular automaton method can tackle the *heterogeneity* of internal and external boundary and starting conditions which is not possible for Avrami-models because they are statistical in nature. The new automaton approach can for instance predict intricate spherulite topologies, kinetic details, and crystallographic textures in homogeneous or heterogeneous fields (the paper gives an example of spherulite growth in a temperature gradient field). Furthermore, it can be coupled to forming and processing models making use of local rather than only global boundary conditions.